\documentclass[fleqn,usenatbib]{mnras}

\usepackage{newtxtext,newtxmath}
\usepackage[T1]{fontenc}
\usepackage{adjustbox}
\usepackage{multirow}
\usepackage{pdflscape}	% Landscape pages
\usepackage{ulem}
\usepackage{mathrsfs}
\DeclareRobustCommand{\VAN}[3]{#2}
\let\VANthebibliography\thebibliography
\def\thebibliography{\DeclareRobustCommand{\VAN}[3]{##3}\VANthebibliography}

\usepackage{graphicx}	% Including figure files
\usepackage{amsmath}	% Advanced maths commands
\usepackage{soul}

\newcommand{\HII}[0]{{}H\,{\sc ii}\,}
\newcommand{\yp}[0]{{}$Y_{\rm p}$}
\title[Primordial Helium Abundance]{A new determination of the primordial helium abundance \\ using the analyses of \HII region spectra from SDSS.}
\author[O.A. Kurichin et al.]{
O.A. Kurichin,$^{1}$\thanks{E-mail: o.chinkuir@gmail.com}
P.A. Kislitsyn,$^{2}$
V.V. Klimenko,$^{1}$
S.A. Balashev,$^{1}$
and A.V. Ivanchik$^{1}$
\\
$^{1}$Ioffe Institute, Polytekhnicheskaya 26, 194021, Saint-Petersburg, Russia\\
$^{2}$Alferov University, Khlopina d.8, k.3, A, 194021, Saint-Petersburg, Russia\\
}

\date{\vspace{-7mm}Accepted 2021 January 21. Received 2020 December 27; in original form 2020 October 01 \\
MNRAS v. 502,  Issue 2, pp. 3045–3056, April 2021}

\pubyear{2021}

\begin{document}
\label{firstpage}
\pagerange{\pageref{firstpage}--\pageref{lastpage}}
\maketitle

\begin{abstract}

The precision measurement of the primordial helium abundance $Y_p$ is a powerful probe of the early Universe. The most common way to determine $Y_p$ is analyses of observations of metal-poor \HII regions found in blue compact dwarf galaxies. We present the spectroscopic sample of 100 \HII regions collected from the Sloan Digital Sky Survey. The final analysed sample consists of our sample and HeBCD database from Izotov et al. 2007. We use a self-consistent procedure to determine physical conditions, current helium abundances, and metallicities of the \HII regions. From a regression to zero metallicity, we have obtained $Y_p = 0.2462 \pm 0.0022$ which is one of the most stringent constraints obtained with such methods up to date and is in a good agreement with the Planck result $Y_{\rm p}^{\it {Planck}} = 0.2471 \pm 0.0003$. Using the determined value of $Y_p$ and the primordial deuterium abundance taken from Particle Data Group (Zyla et al. 2020) we put a constraint on the effective number of neutrino species $N_{\rm eff} = 2.95 \pm 0.16$ which is consistent with the Planck one $N_{\rm eff} = 2.99 \pm 0.17$.  Further increase of statistics potentially allows us to achieve Planck accuracy, which in turn will become a powerful tool for studying the self-consistency of the Standard Cosmological Model and/or physics beyond.

\end{abstract}


%%%%%%%%%%%%%%%%%%%%%%%%%%%%%%%%%%%%%%%%%%%%%%%%%%

%%%%%%%%%%%%%%%%% BODY OF PAPER %%%%%%%%%%%%%%%%%%

\section{Introduction}

The determination of the primordial element abundances is a powerful method for studying the physics of the early Universe. The light  elements D, $^3$He, $^4$He, $^7$Li were produced during Primordial Nucleosynthesis which began a few seconds  after the Big Bang and lasted for several minutes. Comparison of theoretical calculations based on the well-established knowledge of nuclear and particle physics with observations of the light element abundances allows one to obtain an independent estimate of one of the key cosmological parameters -- the baryon-to-photon ratio $\eta \equiv n_b/n_{\gamma}$  \citep[see e.g., ][]{Weinberg2008book,Gorbunov2011book}. Since this quantity is measured independently for the later cosmological epoch (the epoch of Primordial Recombination), it gives us the possibility of checking the Standard Cosmological Model for self-consistency or studying the effects related to the new physics \citep[or ``physics beyond', see e.g. ][]{Cyburt2005}.

Direct measurements of the primordial element abundances are rather complicated by the presence of non-primordial fractions of these elements in the studied objects. This non-primordial fraction mostly arises from stellar nucleosynthesis \citep[e.g.][]{Nomoto2013} thus in order to take this fraction into account, one needs to observe objects with a low star-formation history. The most common way to determine the primordial helium abundance $Y_p$ (the fraction of the primordial helium in the total mass of baryonic matter) is the  analysis of emission lines produced in metal-deficient \HII regions located in blue compact dwarf (BCD) galaxies. Interstellar medium (ISM) of BCDs is chemically relatively unevolved, and thus its elemental composition may be close to the primordial one \citep[see e.g. ][]{1994ApJ...435..647I}. Since the non-primordial fraction of $^4$He is produced over time due to stellar nucleosynthesis as well as the heavier elements, one can expect a correlation between the observed $^4$He abundance $Y$ and the metallicity $Z$ (the abundance of elements heavier than He). Thus the primordial $^4$He abundance \yp\ can be estimated by extrapolation of the  $Y - {\rm Z}$ dependence to zero metallicity. This technique was firstly introduced by \citet{1974ApJ...193..327P} and is still in use \citep{ITG14, AOS15, PPL2016, 2019MNRAS.487.3221F, valerdi2019, hsyu2020}.

\begin{figure}
    \centering
	\includegraphics[width=\columnwidth]{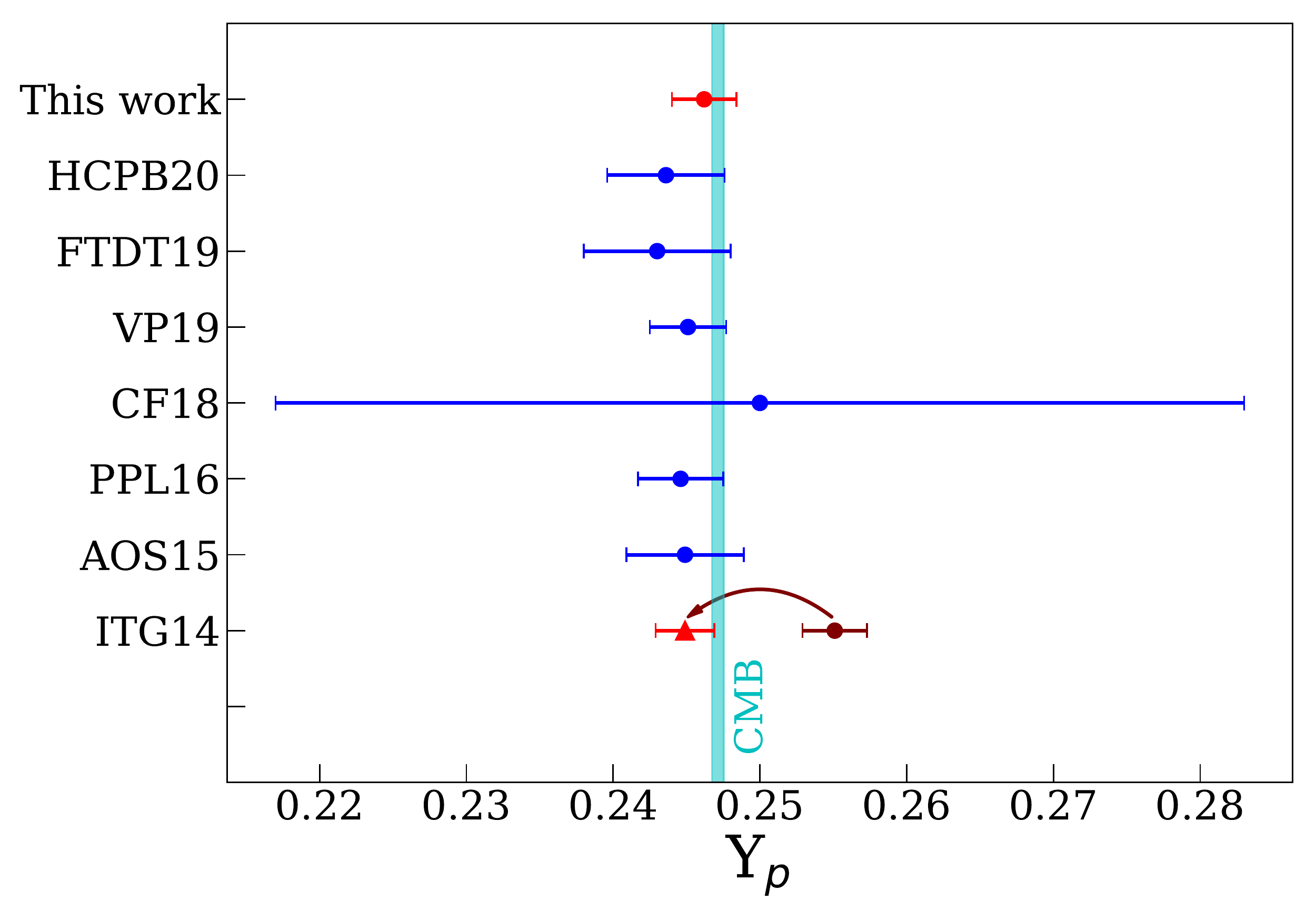}
    \caption{The estimates of \yp\ obtained by different authors and methods (data taken from Table \ref{tab:helium_dets}). The vertical cyan line represents the {\it{Planck}} CMB+BBN prediction for \yp. Our estimate is marked by the red point. We mark the deviating estimate of \protect\cite{ITG14} by the maroon point. The result obtained with the corrected ITG14 procedure (see Sec. \ref{sec:ITG14corr}) is marked by the red triangle point.}
    \label{fig:example_figure}
\end{figure}

According to the \cite{1974ApJ...193..327P}, the helium abundance determination requires measurements of relative fluxes of helium, hydrogen, and metal emission lines. However, there are numerous systematic effects which significantly affect observed fluxes: interstellar reddening, underlying stellar absorption, collisional excitation, self-absorption etc. \citep[e.g.][]{ITG14, AOS15}. In order to account for them the photoionization models for \HII\ regions are required. Currently, there are several independent scientific groups which developed specific models to solve this problem. They obtained similar values of the primordial helium abundances, however there is an inconsistency between the estimation by \cite{ITG14} and others. The reasons for this inconsistency has not yet been found and therefore, for instance, the Particle Data Group \citep[PDG, ][]{10.1093/ptep/ptaa104} reports all mentioned values of $Y_p$ but recommends to use the value $Y_{\rm p} = 0.2450 \pm 0.0030$, which is consistent with the {\it{Planck}} result $Y_{\rm p}^{\it {Planck}} = 0.2471 \pm 0.0003$ (in the

 abstract and the text of the paper we quote the Planck values and their uncertainties at 68\% CL taken from the paper of \cite{planck2018}). We present our and known estimates of \yp\ in Table\,\ref{tab:helium_dets} and Fig.\,\ref{fig:example_figure}.

The largest data sample of metal deficient galaxies was composed by \citet{izotov1997,izotov1998,izotov2004}. The sample contains optical spectra of \HII\ regions in BCD galaxies, and is called the HeBCD database \citep{izotov2007}. For each object in the HeBCD sample, the authors \citep[][ hereafter ITG14]{ITG14} derived physical conditions and heavy-element abundances using so-called ``direct method'' \citep{izotov2006}. The authors used five lines He\,{\sc i} in the optical band: $\lambda 3889$, $\lambda 4471$, $\lambda5876$, $\lambda 6678$, $\lambda 7065$, and one line in the near-infrared (NIR) band: He\,{\sc i} $\lambda 10830$. They found that adding the NIR $\lambda 10830$ line improves the precision of the physical parameter determination and thus reduces the uncertainty of the $Y$ estimate. Using a routine based on Monte-Carlo method \citet{ITG14} presented the final result $Y_{\rm p} = 0.2551\pm0.0022$ which exceeds the value of $Y_{\rm p}^{\it {Planck}} = 0.2471\pm 0.0003$ inferred from the temperature fluctuations of the CMB radiation \citep{planck2018}. This difference may be due to peculiarities of the routine or alternatively may indicate deviations from the Standard Cosmological Model (e.g., the presence of additional relativistic degrees of freedom, dark radiation etc). 

\cite{AOS15} (AOS15) presented independent estimate of $Y_p$ using the same the HeBCD database and NIR observations from ITG14. AOS15 used Markov Chain Monte Carlo method (MCMC) on 8-dimensional parameter space in order to  self-consistently determine the best-fit parameters for the photoionization model of \HII\ regions. AOS15 used 6 lines He\,{\sc i} in the optical range: $\lambda 3889$, $\lambda 4026$, $\lambda 4471$, $\lambda5876$, $\lambda 6678$, $\lambda 7065$, and the near-infrared (NIR) line He\,{\sc i} $ \lambda 10830$. The authors obtained a result of $Y_{\rm p} = 0.2449 \pm 0.0040$. This estimate is consistent with the {\it{Planck}} result but differs form the ITG14 result.

\cite{PPL2016} (PPL16) used a different approach to a determination of $Y_p$. For each object they separately calculated a fraction of $^4$He abundance $\Delta Y$ produced as a result of stellar nucleosynthesis and subtracted this fraction from the observed $Y$. PPL16 presented $Y_{\rm p} = 0.2446 \pm 0.0029$ based on five objects observations. Applying this method to the high-resolution observation of the NGC 346 obtained at the Very Large Telescope (VLT), \cite{valerdi2019} (VP19) presented $Y_{\rm p} = 0.2451 \pm 0.0026$ based on an analysis of the only object. These two results are both consistent with the AOS15 and Planck estimates.

\citet{2019MNRAS.487.3221F} (FTDT19) presented the independent estimation of the \yp\ using their own observational sample. Method proposed by the authors implies a step by step excision of various systematic effects from observed spectra, including the underlying stellar absorption and reddening. The determination of the helium abundance and metallicity of objects was preformed via minimization of the likelihood function calculated for three observed He\,{\sc i} lines - $\lambda 4471$, $\lambda5876$, $\lambda 6678$ (while AOS15 used 6 optical and 1 NIR lines) and a number of metal lines (oxygen, nitrogen, and sulfur). The authors obtained the result: $Y_{\rm p} = 0.2430 \pm 0.0050$.

\cite{2018NatAs...2..957C} proposed a completely different approach to determine \yp. The authors analysed absorption produced by near-pristine intergalactic gas along the line of sight to the quasar HS\,1700$+$5416 (a similar approach is used for the determination of the primordial deuterium abundance). A significant advantage of this approach is that it is not necessary to account for systematic effects of \HII\ region models which might bias the \yp\ estimate. On the other hand the precision of this method is considerably lower (up to date) than the previously discussed approaches. The authors reported $Y_{\rm p} = 0.250 \pm 0.033$.

The most recent estimate of $Y_{\rm p}$ is presented in \cite{hsyu2020}. Their sample consists of the Keck observations, objects selected from Sloan Digital Sky Survey (SDSS) \citep{Aguado2019}, and HeBCD+NIR objects. The authors used a method similar to the AOS15 and obtained $Y_{\rm p} = 0.2436 \pm 0.0040$. 

\begin{table}
	\centering
	\caption{The abundance of the primordial helium \yp\ obtained by different authors and methods.}
	\label{tab:helium_dets}
	\begin{tabular}{lcllr} % four columns, alignment for each
		\hline
		& \hspace{-4mm} $Y_p$ & & Paper & Abbreviation\\
		\hline
		0.2551  & \hspace{-4mm}$\pm$ & \hspace{-4mm} 0.0022  & \citet{ITG14} & ITG14\\
	    0.2449  & \hspace{-4mm}$\pm$ & \hspace{-4mm} 0.0040  & \citet{AOS15}& AOS15\\
		0.2446  & \hspace{-4mm}$\pm$ & \hspace{-4mm} 0.0029  &  \citet{PPL2016}& PPL16\\
		0.250  & \hspace{-4mm}$\pm$ & \hspace{-4mm} 0.0330   &\cite{2018NatAs...2..957C}& CF18\\
		0.243   & \hspace{-4mm}$\pm$ & \hspace{-4mm} 0.0050   & \citet{2019MNRAS.487.3221F}& FTDT19 \\
		0.2451  & \hspace{-4mm}$\pm$ & \hspace{-4mm} 0.0026  & \citet{valerdi2019} & VP19\\
		0.2436 & \hspace{-4mm}$\pm$ & \hspace{-4mm} 0.0040   &\cite{hsyu2020}& HCPB20\\
		{\it 0.2462}   & \hspace{-4mm}$\pm$ & \hspace{-4mm} {\it 0.0022}   & This paper &\\
		\hline
        0.2471 & \hspace{-4mm} $\pm$ & \hspace{-4mm} 0.0003 & \citet{planck2018} &\\
		\hline
	\end{tabular}
\end{table}

In this paper we present the systematic study of the helium abundance in \HII\ regions located in BCD galaxies. We reproduce the methods for the \yp\ estimation proposed by \citet{ITG14} and \citet{AOS15} and also present our independent result for \yp \, based on analyses of SDSS spectra of BCD galaxies, such a possibility was discussed in our previous work \citep{2019JPhCS1400b2051K} .  The structure of the paper is as follows. The sample of \HII\ regions selected from the SDSS catalog is presented in Sect.\,\ref{sec:2sdss}. In Sect.\,\ref{sec:3} we present our estimates of the primordial helium abundance and slope of $Y - {\rm O/H}$ relation. In Sect.\,\ref{sec:4} we discuss obtained results, reasons for discrepancy with \cite{ITG14} and further possible improvements, before we conclude in Sect.\,\ref{sec:concl}. In this paper we quote 68\% confidence regions on measured parameter.

\section{Data reduction and the sample compilation}
\label{sec:2sdss}

In this section, we describe observational sample of BCD galaxies. The spectra are selected from the SDSS catalog. We determine the physical properties of \HII\ regions and the abundance of helium  using the approach similar to AOS15 (see Appendix A for details). 

\subsection{The SDSS sample}
The determination of the primordial helium abundance requires the analyses of high quality \HII region spectra. Despite the medium resolution and signal-to-noise (S/N) ratio of the SDSS spectra, a great amount of SDSS objects can drastically increase statistics. We use the spectroscopic data from the SDSS Data Release 15 (DR15, \citealt{Aguado2019}) which contains the flux-calibrated spectra of 80\,420 starburst galaxies (this objects are tagged as 'STARBURST' in the SDSS catalog). We scanned the SDSS catalogs in a search for metal-poor \HII regions. From the catalog we selected spectra of \HII regions satisfying the following selection criteria: (1) the redshift is within the range of $0 \leq z \leq 0.3$ so that the longest wavelength line used in our analysis (He$\lambda7065$) to be within the spectrograph band of 3800-9200\,\AA\, it gives 71\,433 objects, (2) the mean S/N $\geq 10$, which leaves 46\,197 objects. Up to date we manually selected and analysed 580 objects (hereafter the $S_0$ sample). 

\begin{figure*}
    \centering
    \includegraphics[width = 0.8\textwidth]{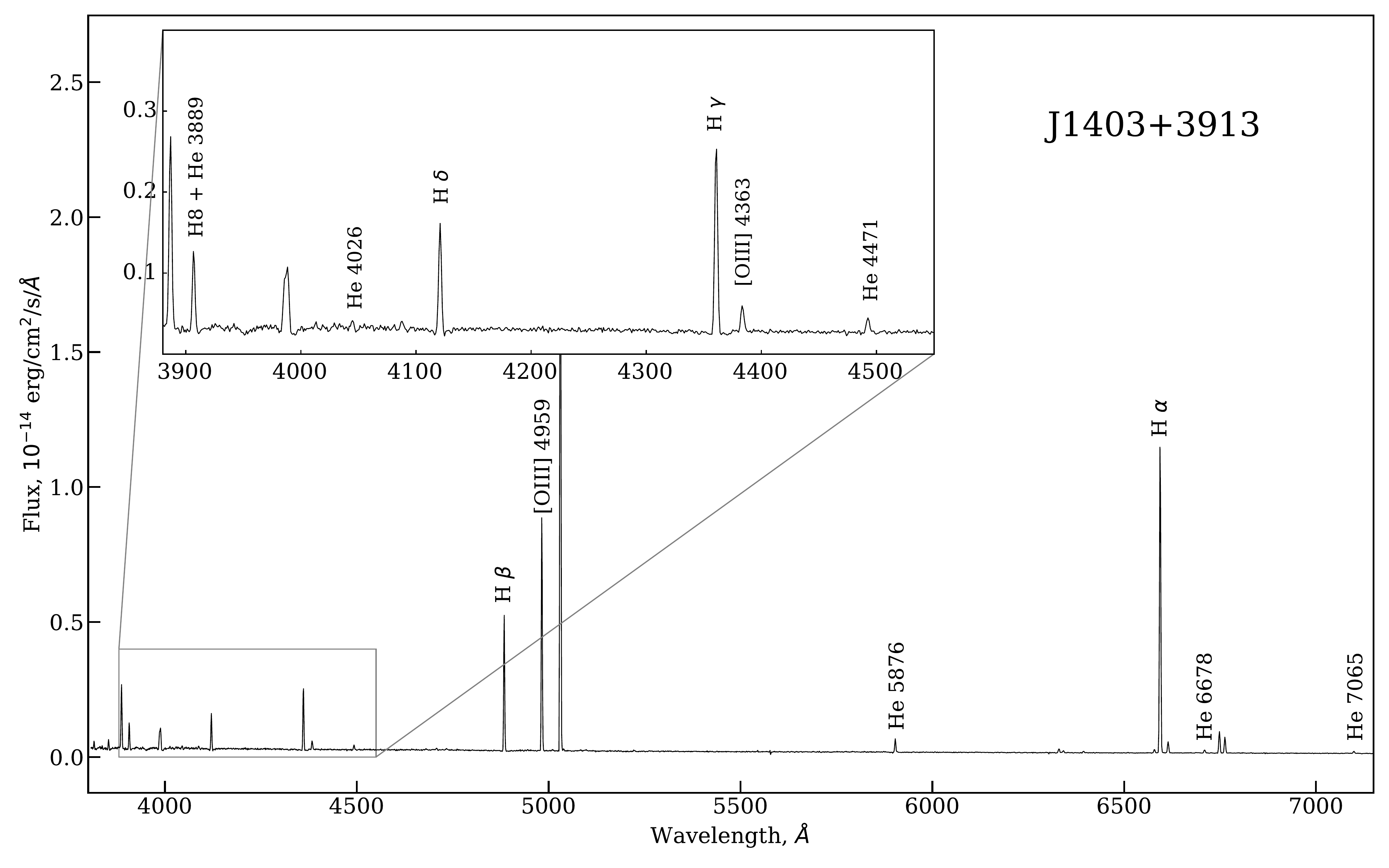}
    \caption{An example of an \HII\ region spectrum from the sample $S_0$. The spectrum has S/N ratio of 22.42 and the metallicity of 12 + $\log$(O/H) = 7.93.}
    \label{fig:typSDSS}
    
\end{figure*}

\subsection{The spectral analysis}
We estimate \yp\ using approach, similar to AOS15. We measure the fluxes and equivalent widths of the following emission lines: He\,{\sc i} ($\lambda3889$, $\lambda4026$, $\lambda4471$, $\lambda5876$, $\lambda6678$, $\lambda7065$), He\,{\sc ii}($\lambda4686$), H\,{\sc i} ($\lambda4101$, $\lambda4340$, $\lambda4861$, $\lambda6563$), metal lines: O\,{\sc iii} ($\lambda4363$, $\lambda4959$, $\lambda5007$),  O\,{\sc ii} ($\lambda7320$, $\lambda7330$), and S\,{\sc ii} ($\lambda6717$ and $\lambda6731$). The spectrum of a typical object from the $S_0$ sample is presented in Fig. \ref{fig:typSDSS}. For each emission line we construct a local continuum by fitting a spline to selected continuum regions devoiding of any absorption and emission. We determine a flux and equivalent width of a line by the Gaussian profile fitting. We fit the profile using Monte Carlo Markov Chain approach with affine invariant sampler to obtain the posterior probability function. The values of fitted parameters (amplitude, centroid and variance) corresponds to the maximum posterior probability and uncertainties estimate corresponds to that containing 68.3\% of area.

\subsection{Measurements of helium abundance and metallicity}

The photoionization model used for the determination of the observed helium mass fraction $Y$ and metallicity is described in Appendix A. Here we present a brief description of its main properties.

The fluxes of hydrogen and helium emission lines are calculated as functions of specific physical parameters and then compared with the measured fluxes. These parameters include: the ratio of number densities of the single ionized helium to hydrogen $y^+$, electron density $n_e$, electron temperature $T_e$, optical depth $\tau$, parameters of He and H underlying absorption $a_{\rm He}$ and $a_{\rm H}$, extinction coefficient C(H$\beta$), and neutral hydrogen fraction $\xi$. Each of the parameters is related to the corresponding systematic effect which changes the intrinsic line fluxes (see Appendix A for details). 

We obtain the probability distribution functions of these parameters with the Monte Carlo Markov Chain approach implementing the affine-invariant ensemble sampler \cite{2019JOSS....4.1864F}. Such a technique ensures that we confidently find the global maximum in the many parametric space and provides reliable estimates of statistical errors on the parameters. Furthermore, the correlation between  different parameters can be easily traced by this method.

\subsection{The final sample}
\label{sec:finalsample}

Applying the procedure described above to the sample $S_0$ and using additional selection criteria, we form the final sample $S_{\rm{f}}$ to be used for a determination of $Y_p$. Firstly, we exclude 72 spectra where He $\lambda 4026$ line was not detected. This line is crucial for $a_{\rm He}$ determination, which in turn allows to  significantly decrease the systematic uncertainty of the $y^+$. It leaves us 508 spectra from the sample $S_0$. We exclude 11 additional spectra, where the weak \ion{O}{iii} $\lambda 4363$ line (which is necessary for a determination of O/H) was not detected. This leaves us with 497 spectra.

Secondly, we select spectra which are well described with the photoionization model, it is characterized with the $\chi^2$ criteria. We assumed that $\chi^2$ value for a good-fit model should be in the range $[\nu - \sqrt{2\nu}, \nu + \sqrt{2\nu}]$, where $\nu$ is the number of degrees of freedom (see e.g. \citealt{hogg}). We selected 100 objects satisfying the $\chi^2$ criteria and form our final sample $S_{\rm f}$ for the regression analysis. For each object in the sample $S_{\rm f}$ the metallicity O/H, $Y$, and $\chi^2$ values are presented in Table \ref{tab:result_table}.

\begin{table*}
	\centering
	\caption{The derived parameters: metallicity O/H, current helium abundance $Y$, and $\chi^2$ value for the objects of the sample $S_{\rm f}$. }
	\label{tab:result_table}
	\begin{tabular}{ccccc@{\qquad}ccccc} % four columns, alignment for each
		\hline
		No. & Object & O/H$\cdot 10^{5}$ & Y & $\chi^2$ & No. & Object & O/H$\cdot 10^{5}$ & Y & $\chi^2$ \\
		\hline
1 & J0024+1404 & 27.44 $\pm$ 4.04 & 0.2633 $\pm$ 0.0063 & 3.78 & 51 & J1202+6220 & 26.94 $\pm$ 2.48 & 0.2496 $\pm$ 0.0127 & 2.43 \\
2 & J0059+0100 & 15.70 $\pm$ 5.26  &0.2572 $\pm$ 0.0116 & 3.19 & 52 & J1207+2507 & 18.18 $\pm$ 2.26 & 0.2502 $\pm$ 0.0088 & 3.53 \\
3 & J0212+0113 & 11.20 $\pm$ 1.31 & 0.2482 $\pm$ 0.0158 & 2.35 & 53 & J1208+5525 & 8.23 $\pm$ 0.51 & 0.2589 $\pm$ 0.0067 & 3.79 \\
4 & J0254-0041 & 11.35 $\pm$ 1.22 & 0.2465 $\pm$ 0.0133 & 3.75 & 54 & J1211+3945 & 18.93 $\pm$ 1.48  &0.2502 $\pm$ 0.0135 & 3.72 \\
5 & J0729+3950 &17.21 $\pm$ 1.32 &0.2664 $\pm$ 0.0079 & 3.97 & 55 & J1212+0004 & 14.42 $\pm$ 1.10  &0.2688 $\pm$ 0.0109 & 3.04 \\
6 & J0806+1949 &19.42 $\pm$ 1.70 &0.2618 $\pm$ 0.0059 & 3.36 & 56 & J1221+0245 & 12.68 $\pm$ 1.43 & 0.2499 $\pm$ 0.0196 & 3.62 \\
7 & J0817+5202 & 19.10 $\pm$ 2.42 & 0.2541 $\pm$ 0.0080 & 3.54 & 57 & J1224+6726 & 27.12 $\pm$ 7.26  &0.2531 $\pm$ 0.0082 & 2.86 \\
8 & J0819+5000 & 25.62 $\pm$ 3.43 & 0.2490 $\pm$ 0.0074 & 2.16 & 58 & J1227+5139 & 18.41 $\pm$ 2.46 & 0.2561 $\pm$ 0.0148 & 3.28   \\
9 & J0826+4558 & 12.98 $\pm$ 1.08  &0.2470 $\pm$ 0.0085 & 3.87 & 59 & J1231+0553 & 13.29 $\pm$ 2.66 & 0.2435 $\pm$ 0.0338 & 3.09 \\
10 & J0827+4602 & 19.60 $\pm$ 3.09  &0.2494 $\pm$ 0.0084 & 3.97 & 60 & J1237+0230 & 13.04 $\pm$ 1.33 & 0.2506 $\pm$ 0.0215 & 2.77 \\
11 & J0830+1427 & 12.03 $\pm$ 0.95 & 0.2443 $\pm$ 0.0073 & 2.80 & 61 & J1238+1010 & 10.84 $\pm$ 0.39 & 0.2543 $\pm$ 0.0091 & 3.55  \\
12 & J0844+0226 & 18.45 $\pm$ 0.76 & 0.2562 $\pm$ 0.0048 & 3.46 & 62 & J1242+3232 & 17.61 $\pm$ 1.96 & 0.2391 $\pm$ 0.0061 & 2.06 \\
13 & J0900+3543 & 16.85 $\pm$ 1.15 & 0.2432 $\pm$ 0.0044 & 3.63 & 63 & J1247+2634 & 25.58 $\pm$ 3.93  &0.2612 $\pm$ 0.0137 & 2.01 \\
14 & J0907+3857 & 23.35 $\pm$ 3.45 &  0.2510 $\pm$ 0.0073 & 2.33 & 64 & J1249+4743 & 14.42 $\pm$ 0.62  & 0.2429 $\pm$ 0.0087 & 2.34 \\
15 & J0907+5327 & 15.26 $\pm$ 1.56 & 0.2436 $\pm$ 0.0066 & 3.28 & 65 & J1301+1240 &21.68 $\pm$ 1.46 &0.2737 $\pm$ 0.0059 & 3.99 \\
16 & J0911+5228 & 15.85 $\pm$ 1.83 & 0.2515 $\pm$ 0.0087 & 2.38 & 66 & J1313+1229 &11.55 $\pm$ 1.46 &0.2501 $\pm$ 0.0059 & 2.37 \\
17 & J0916+4300 & 15.16 $\pm$ 1.00 &0.2668 $\pm$ 0.0167 & 2.35 & 67 & J1314+3909 & 10.97 $\pm$ 0.69  &0.2445 $\pm$ 0.0074 & 3.95 \\
18 & J0928+3808 & 14.97 $\pm$ 1.17 & 0.2474 $\pm$ 0.0104 & 2.32  & 68 & J1330+3120 & 18.30 $\pm$ 1.14  &0.2686 $\pm$ 0.0162 & 2.84 \\
19 & J0936+3130 & 14.46 $\pm$ 3.04 & 0.2420 $\pm$ 0.0165 & 3.13 &69 & J1343+5242 & 12.26 $\pm$ 0.89 & 0.2475 $\pm$ 0.0092 & 2.72  \\
20 & J0946+0140 & 4.72 $\pm$ 2.48 & 0.2548 $\pm$ 0.0163 & 3.43 & 70 & J1344+5601 & 20.84 $\pm$ 1.60 & 0.2611 $\pm$ 0.0155 & 3.89  \\
21 & J0947+5406 & 4.56 $\pm$ 1.12 & 0.2442 $\pm$ 0.0173 & 1.11 & 71 & J1347+6202 & 12.56 $\pm$ 1.41 & 0.2490 $\pm$ 0.0156 & 3.20 \\ 
22 & J0948+4257 & 19.41 $\pm$ 2.69 & 0.2603 $\pm$ 0.0071 & 1.72  & 72 & J1402+3913 & 11.34 $\pm$ 0.71  &0.2483 $\pm$ 0.0082 & 0.92 \\
23 & J0949+0143 & 18.15 $\pm$ 2.27 & 0.2598 $\pm$ 0.0079 & 1.44 &73 & J1402+5416 & 19.51 $\pm$ 3.78  &0.2430 $\pm$ 0.0051 & 3.74 \\
24 & J0952+0218 & 15.82 $\pm$ 2.24 & 0.2682 $\pm$ 0.0182 & 3.82 & 74 & J1403+3913 & 11.37 $\pm$ 0.79  & 0.2520 $\pm$ 0.0075 & 0.98  \\
25 & J0957+3337 & 25.78 $\pm$ 3.09  &0.2676 $\pm$ 0.0089 & 2.34 & 75 & J1404+5424 & 12.44 $\pm$ 0.41 & 0.2707 $\pm$ 0.0127 & 3.11  \\
26 & J1000+3833 & 18.60 $\pm$ 4.13 & 0.2682 $\pm$ 0.0108 & 2.92  & 76 & J1413+1830 & 9.75 $\pm$ 0.35 &0.2470 $\pm$ 0.0104 & 3.62 \\
27 & J1001+0112 & 5.21 $\pm$ 0.70 &0.2364 $\pm$ 0.0156 & 3.30 & 77 & J1418+1935 & 21.39 $\pm$ 3.02 & 0.2546 $\pm$ 0.0097 & 2.14 \\
28 & J1002+3715 & 17.79 $\pm$ 0.92  & 0.2626 $\pm$ 0.0077 & 3.83 & 78 & J1425+5133 & 17.47 $\pm$ 1.66 & 0.2496 $\pm$ 0.0090 & 2.92  \\
29 & J1006+2857 & 10.47 $\pm$ 2.65  &0.2419 $\pm$ 0.0172 & 1.82 & 79 & J1426+6300 & 16.13 $\pm$ 1.06 & 0.2434 $\pm$ 0.0089 & 3.34 \\
30 & J1012+1221 &13.74 $\pm$ 0.45 &0.2504 $\pm$ 0.0078 & 3.82 & 80 & J1428+1333 & 8.72 $\pm$ 1.19 & 0.2520 $\pm$ 0.0072 & 2.32 \\
31 & J1015+5951 &5.56 $\pm$ 0.86 &0.2377 $\pm$ 0.0150 & 3.81 & 81 & J1430+0643 &15.17 $\pm$ 0.67 &0.2512 $\pm$ 0.0085 & 2.57 \\
32 & J1032+2325 & 14.17 $\pm$ 1.10  &0.2409 $\pm$ 0.0068 & 3.15 & 82 & J1430+4532 & 23.57 $\pm$ 2.47  &0.2751 $\pm$ 0.0108 & 3.08 \\
33 & J1041+0625 & 19.71 $\pm$ 3.73 & 0.2579 $\pm$ 0.0159 & 2.58 & 83 & J1432+3141 &8.92 $\pm$ 0.80 &0.2518 $\pm$ 0.0069 & 3.70 \\
34 & J1041+2122 & 11.41 $\pm$ 0.45  &0.2463 $\pm$ 0.0133 & 2.44 & 84 & J1432+5153 & 11.77 $\pm$ 0.80 & 0.2411 $\pm$ 0.0124 & 3.37 \\
35 & J1046+0104 & 16.00 $\pm$ 0.30 & 0.2575 $\pm$ 0.0096 & 2.34 & 85 & J1433+0255 & 14.30 $\pm$ 3.79 & 0.2737 $\pm$ 0.0241 & 3.63 \\
36 & J1048+1308 & 13.90 $\pm$ 1.71 & 0.2565 $\pm$ 0.0174 & 3.32 & 86 & J1437+0303 & 9.80 $\pm$ 0.90 & 0.2551 $\pm$ 0.0153 & 3.48 \\ 
37 & J1049+2603 & 22.81 $\pm$ 1.68 & 0.2679 $\pm$ 0.0049 & 2.16 & 87 & J1441+0041 & 17.18 $\pm$ 1.96 & 0.2541 $\pm$ 0.0128 & 3.45 \\
38 & J1049+3259 & 29.23 $\pm$ 10.22  &0.2534 $\pm$ 0.0055 & 1.77 & 88 & J1443+2818 & 17.93 $\pm$ 3.58 & 0.2582 $\pm$ 0.0134 & 1.51 \\
39 & J1059+5929 & 24.73 $\pm$ 6.62 & 0.2480 $\pm$ 0.0061 & 2.89  & 89 & J1447+1246 & 14.70 $\pm$ 1.64  &0.2659 $\pm$ 0.0105 & 3.71 \\
40 & J1118+1851 & 24.56 $\pm$ 4.80  &0.2587 $\pm$ 0.0037 & 2.15 & 90 & J1457+3014 &11.82 $\pm$ 0.46 & 0.2726 $\pm$ 0.0083 & 2.84 \\
41 & J1120+4728 & 16.68 $\pm$ 3.26 & 0.2628 $\pm$ 0.0248 & 3.21 & 91 & J1510+3732 & 7.49 $\pm$ 0.21  & 0.2426 $\pm$ 0.0099 & 3.66 \\
42 & J1125+5716 & 13.18 $\pm$ 3.48  &0.2596 $\pm$ 0.0280 & 3.37 & 92 & J1517-0008 & 14.69 $\pm$ 1.90 & 0.2462 $\pm$ 0.0079 & 0.87 \\
43 & J1126-0040 & 15.20 $\pm$ 3.52 & 0.2440 $\pm$ 0.0118 & 3.24 & 93 & J1538+4548 & 14.28 $\pm$ 2.11 & 0.2702 $\pm$ 0.0244 & 2.85 \\
44 & J1128+2047 & 13.18 $\pm$ 1.10 & 0.2548 $\pm$ 0.0121 & 3.98 & 94 & J1557+2321 &21.63 $\pm$ 1.44 &0.2583 $\pm$ 0.0054 & 2.91 \\
45 & J1137+3624 & 17.57 $\pm$ 1.51  &0.2428 $\pm$ 0.0067 & 2.97 & 95 & J1608+4940 & 15.75 $\pm$ 2.64 & 0.2553 $\pm$ 0.0080 & 1.61 \\ 
46 & J1143+3127 & 13.56 $\pm$ 2.69   &0.2689 $\pm$ 0.0082 & 3.63 & 96 & J1632+1338 & 26.61 $\pm$ 3.62 & 0.2629 $\pm$ 0.0087 & 3.63 \\
47 & J1143+6807 & 11.06 $\pm$ 1.12 & 0.2499 $\pm$ 0.0105 & 1.07 & 97 & J1642+4223 & 15.78 $\pm$ 2.04 & 0.2471 $\pm$ 0.0109 & 0.99 \\ 
48 & J1149+3502 &22.21 $\pm$ 2.57 &0.2512 $\pm$ 0.0047 & 2.62 & 98 & J1644+4436 & 24.56 $\pm$ 6.33 & 0.2638 $\pm$ 0.0123 & 3.05 \\ 
49 & J1150+2409 & 13.68 $\pm$ 3.00 & 0.2671 $\pm$ 0.0161 & 2.85 & 99 & J2115-0800 & 20.30 $\pm$ 1.99 & 0.2486 $\pm$ 0.0123 & 2.54 \\ 
50 & J1151+4815 & 11.60 $\pm$ 1.03  & 0.2444 $\pm$ 0.0095 & 2.77 & 100 & J2320-0052 & 13.44 $\pm$ 1.67 & 0.2592 $\pm$ 0.0175 & 2.19 \\
    \hline
	\end{tabular}
	\hspace{1em}

\end{table*}

\subsection{The HeBCD subsample}
\label{sec:HEBCDsample}
In addition to our final sample $S_{\rm f}$ (Table \ref{tab:result_table}) we use the HeBCD database, which is one of the largest databases of high quality spectra of metal deficient \HII regions. We use the optical spectra from \cite{izotov2007} and the NIR spectra from \cite{ITG14} and determine the physical properties and abundances of helium and oxygen of the HeBCD objects using the approach discussed above. Applying the selection criteria from Section \ref{sec:finalsample} to HeBCD database we selected 20 objects (for comparison AOS15 selected 17 objects and ITG14 selected 28). We present $Y$, O/H and $\chi^2$ for these objects in Table \ref{tab:HeBCD_result_table}. Note that, it could be seen from Fig. \ref{fig:fullregr}, Tab. \ref{tab:result_table} and \ref{tab:HeBCD_result_table} that HeBCD objects have less uncertainty in O/H than SDSS objects, and at the same time their uncertainties in $Y$ are comparable. The large uncertainty in O/H for SDSS objects is associated with lower S/N ratio of their spectra. The fact is that the uncertainty in O/H is directly determined by the error in the measured flux of the weak line [\ion{O}{iii}] $\lambda$4363, therefore the lower S/N ratio means the larger uncertainty in O/H. The uncertainty in $Y$ for the SDSS and HeBCD objects is comparable, since it is determined by the photoionization model itself and weakly depends on the errors in fluxes of individual He lines.

\begin{table}
	\centering
	\caption{The derived parameters: metallicity O/H, current helium abundance $Y$, and $\chi^2$ value for the objects of the HeBCD sample.}
	\label{tab:HeBCD_result_table}
	\begin{tabular}{ccccc} % four columns, alignment for each
		\hline
		No. & Object & O/H$\cdot 10^{5}$ & Y & $\chi^2$ \\
		\hline
		1 & CGCG 007--025 & 5.5 $\pm$ 0.2 & 0.2542 $\pm$ 0.0210 & 2.31  \\
2 & I Zw 18 SE & 1.5 $\pm$ 0.1 & 0.2366 $\pm$ 0.0137 & 0.95\\
3 & J0519+0007 & 2.8 $\pm$ 0.1 & 0.2689 $\pm$ 0.0120 & 5.41 \\
4 & Mrk 1315  & 18.9 $\pm$ 0.4 & 0.2618 $\pm$ 0.0096 & 5.14  \\
5 & Mrk 1329  & 19.2 $\pm$ 0.5 & 0.2692 $\pm$ 0.0158 & 1.54 \\
6 & Mrk 209 & 6.1 $\pm$ 0.1 & 0.2484 $\pm$ 0.0026 & 5.62\\
7 & Mrk 450 № 1 & 15.2 $\pm$ 0.4 & 0.2546 $\pm$ 0.0063 & 5.68 \\
8 & Mrk 59 & 10.1 $\pm$ 0.2 & 0.2549 $\pm$ 0.0232 & 4.04 \\ 
9 & Mrk 71 & 7.2 $\pm$ 0.2 & 0.2542 $\pm$ 0.0128 & 4.66\\ 
10 & SBS 0335--052E & 2.0 $\pm$ 0.1 & 0.2497 $\pm$ 0.0069 & 4.58  \\
11 & SBS 0940+544 & 3.2 $\pm$ 0.1 & 0.2471 $\pm$ 0.0068 & 2.41 \\
12 & SBS 1030+583 & 6.4 $\pm$ 0.2 & 0.2470 $\pm$ 0.0159 & 1.67 \\
13 & SBS 1135+581 & 11.7 $\pm$ 0.3 & 0.2535 $\pm$ 0.0147 & 4.41 \\
14 & SBS 1152+579 & 7.7 $\pm$ 0.2 & 0.2473 $\pm$ 0.0064 & 4.91 \\
15 & SBS 1222+614 & 9.8 $\pm$ 0.2 & 0.2462 $\pm$ 0.0197 & 2.76 \\
16 & SBS 1415+437 № 1 & 4.0 $\pm$ 0.1 & 0.2415 $\pm$ 0.0125 & 4.96 \\
17 & SBS 1415+437 № 2 & 4.2 $\pm$ 0.3 & 0.2488 $\pm$ 0.0096 & 2.45\\
18 & Tol 1214--277 & 3.5 $\pm$ 0.1 & 0.2596 $\pm$ 0.0105 & 3.25 \\
19 & Tol 65 & 3.5 $\pm$ 0.1 & 0.2431 $\pm$ 0.0105 & 5.12\\
20 & UM 311 & 20.4 $\pm$ 0.2 & 0.2464 $\pm$ 0.0089 & 1.80  \\
    \hline	
	\end{tabular}
	\hspace{1em}
\end{table}

\section{REGRESSION ANALYSIS}
\label{sec:3}
\begin{figure*}
\centering
	\includegraphics[scale = 0.35]{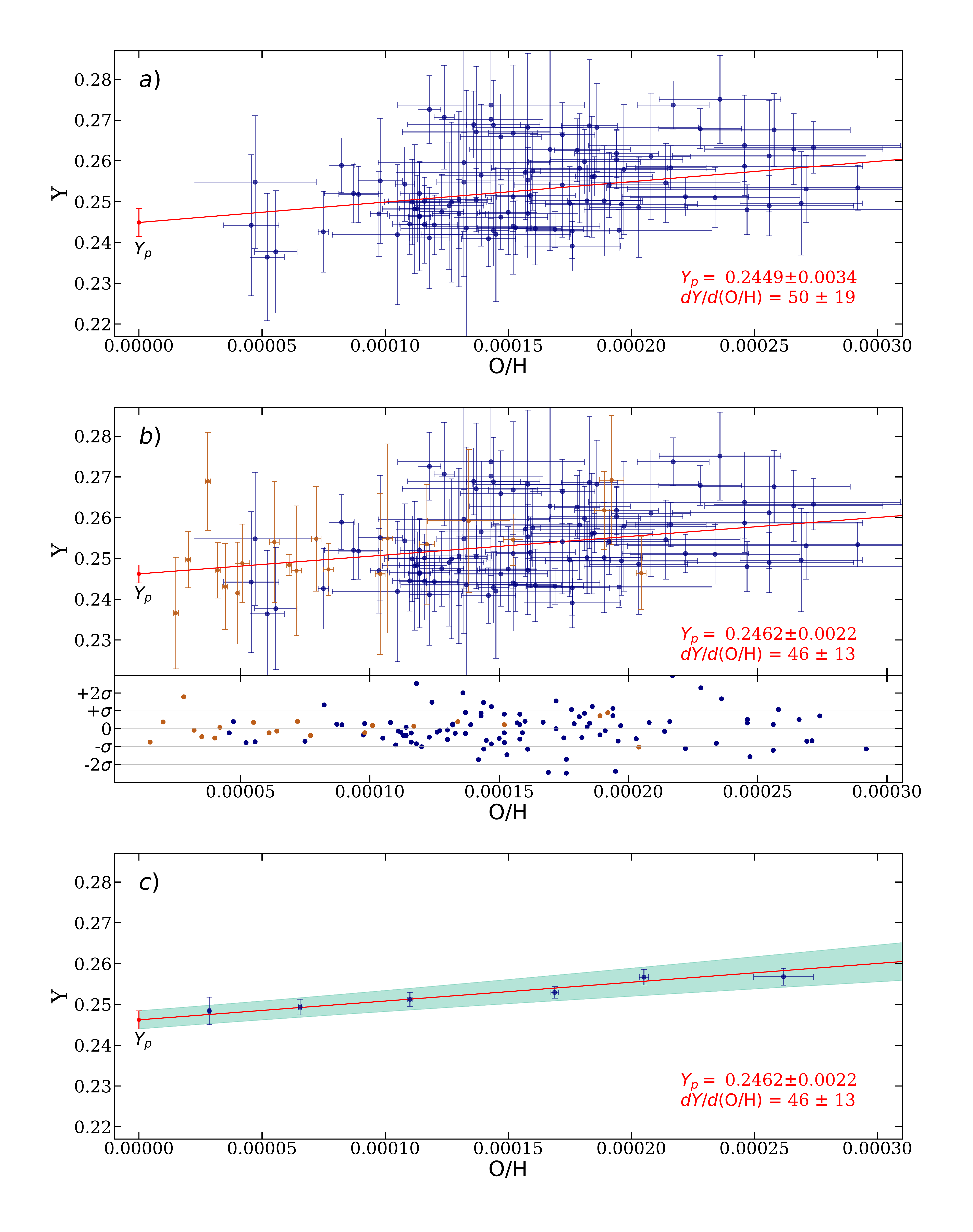}
    \caption{{\textbf{(a)}} ``$Y$ - (O/H)'' diagram for 100 \HII regions of the final sample $S_{\rm f}$. {\textbf{(b)}} ``$Y$ - (O/H)'' diagram for the final sample $S_{\rm f}$ (navy colored) combined with 20 \HII regions of the HeBCD subsapmle (brown colored). The panel beneath the plot {\textbf{(b)}} represents the dispersion of points around the regression line in terms of sigma intervals with $\sim$75\% fraction of all points falling into 1$\sigma$ interval and $\sim$94\% falling into 2$\sigma$ interval. {\textbf{(c)}} the same as {\textbf{(b)}} but the whole sample was sliced into equally-sized bins in which of those the weighted mean point was calculated for a grater visualisation and clarity (regression line on {\textbf{(c)}} is the same as on {\textbf{(b)}}). }
    \label{fig:fullregr}
\end{figure*}

The He abundances $Y$ derived from the emission line analysis in metal-poor starburst galaxies of the $S_{\rm f}$ sample are presented in Fig.\ref{fig:fullregr} together with the measurements obtained from the HeBCD subsample.

\begin{table}
	\centering
	\caption{Estimates of the \yp\ and $dY/d(O/H)$.}
	\label{tab:ourestimate}
	\begin{tabular}{ccc} % four columns, alignment for each
		\hline
		Sample & \yp\ & $dY$/$d$(O/H) \\
		\hline
		HeBCD & $0.2464\pm0.0030$ & $51\pm31$ \\
		$S_{\rm f}$ & $0.2449\pm0.0034$ & $50\pm19$ \\
		$S_{\rm f}$ + HeBCD & $0.2462\pm0.0022$ & $46\pm13$ \\
		\hline
	\end{tabular}
\end{table}

We perform a regression analysis of $Y$ versus O/H values with the following expression: 
\begin{equation}
    \label{eq_fit}
    Y = Y_p + \frac{dY}{d({\rm O/H})}\times({\rm O/H})
\end{equation}

We estimate \yp\ and the slope $dY/d({\rm O/H})$ with the MCMC routine in the following cases: (i) for the $S_{\rm f}$ and HeBCD samples separately, and (ii) their combination. The results are summarized in Table\,\ref{tab:ourestimate}. One can note that the fit to the SDSS and HeBCD samples gives similar results. However, larger number of SDSS objects and a wider range of metallicities improve the precision of the slope estimate by a factor of about 1.5. Combining two samples gives us the final estimate:
\begin{equation}
    Y_p = 0.2462 \pm 0.0022 \text{ ~~~ and ~~~~} dY/d(O/H) = 46 \pm 13
\end{equation}
The obtained $Y_p$ is in a good agreement with Planck's result $Y^{Planck}_p = 0.2471 \pm 0.0003$ (\citealt{planck2018}).

Fig. \ref{fig:fullregr}{\textbf{(c)}} is a simplified visualisation of Fig. \ref{fig:fullregr}{\textbf{(b)}}. The entire range of metalicities is divided into equal bins for which the weighted mean of the points was calculated. The means and their uncertainties are plotted on a $Y-$(O/H) plane along with a regression curve, 1 $\sigma$ interval and $Y_p$ which were obtained earlier. The figure shows that the linear correlation of $Y- $(O/H) is preserved at least up to O/H $\sim 25 \times 10^{-5}$. 

Theoretical and observational estimates of the increase of helium with the increase of oxygen were discussed in \cite{2007ASPC..374...81P}. Note that the authors use the oxygen abundance O by mass fraction instead of the number density abundance O/H (which we use in this paper). These quantities are connected by the relation O = $\frac{16 \cdot \rm{O/H}}{1 + 4y}$, where $y$ is the helium number density. The authors showed that the linear relation $Y-$O is preserved up to O $\sim 4 \times 10^{-3}$, which in our terms corresponds to O/H $\sim 30 \times 10^{-5}$. Additionally, the authors derived the slope of the relation $Y-$O. Rewriting our estimate of the slope $dY/d({\rm O/H}) = 46 \pm 13$ in Peimbert's terms, we get the value $\Delta$Y/$\Delta$O = $3.4 \pm 1.1$, which is in a good agreement with the value $\Delta$Y/$\Delta$O = $3.3 \pm 0.7$ presented in \cite{2007ASPC..374...81P}.

\subsection{Estimation of \texorpdfstring{$N_{\rm eff}$}{Lg}}

In the frame of Primordial Nucleosynthesis the primordial $^4$He abundance $Y_p$ has a strong dependence on the effective number of neutrino species $N_{\rm eff}$ which allows to constrain this quantity using the estimated $Y_p$ value. To constrain $N_{\rm eff}$ we use the following relation from \cite{2020JCAP...03..010F}:
\begin{equation}
\begin{split}
     &Y_p = 0.24696 \Big ( \frac{\eta_{10}}{6.129}\Big) ^{0.039} \Big ( \frac{N_{\rm eff}}{3.0}\Big) ^{0.396} \Big ( \frac{G_{\rm N}}{G_{\rm N, 0}}\Big) ^{0.952} \Big ( \frac{\tau_n}{879.4}\Big) ^{0.409} \\
     &\times [p(n,\gamma)d]^{0.005} [d(d,n)^3He]^{0.006} [d(d,p)t]^{0.005}
\end{split}
\label{eq:CFOY}
\end{equation}

Here $\eta_{10} = 10^{10} \eta$, $G_{\rm N}$ is Newtonian gravitational constant, $\tau_n$ is a neutron mean lifetime. The last three terms are the key nuclear rates which affect the $Y_p$. Note that $Y_p$ is not a precise baryometer, but instead, $Y_p$ sets a tight constraint on the effective number of neutrino species. To estimate $\eta_{10}$ we use a precise measurement of the primordial deuterium abundance $D/H = (2.538 \pm 0.025) \times 10^{-5}$ from Particle Data Group \citep{10.1093/ptep/ptaa104} which is a weighted mean of 16 measurements. To constrain $\eta_{10}$ we use the following relation from \cite{2020JCAP...03..010F}:

\vspace{4mm}

\begin{equation}
\begin{split}
     &\frac{D}{H} = 2.559 \times 10^{-5} \Big ( \frac{\eta_{10}}{6.129}\Big) ^{-1.597} \Big ( \frac{N_{\rm eff}}{3.0}\Big) ^{0.163} \Big ( \frac{G_{\rm N}}{G_{\rm N, 0}}\Big) ^{0.357}  \\
     &\times \Big ( \frac{\tau_n}{879.4}\Big) ^{0.729} [p(n,\gamma)d]^{-0.193} [d(d,n)^3He]^{-0.529} [d(d,p)t]^{-0.047} \\
     &\times [d(p,\gamma)^3He]^{-0.315} [^3He(n, p)t]^{0.023} [^3He(d, p) ^4He]^{-0.012}
\end{split}
\label{eq:CFOYd}
\end{equation}

Using MCMC procedure we get the following estimates of $\eta_{10}$ and $N_{\rm eff}$: $\eta_{10} = 6.14 \pm 0.09$  and $N_{\rm eff} = 2.95 \pm 0.16$. This estimates are in a good agreement with the Planck results \citep{planck2018} $\eta_{10} = 6.13 \pm 0.04$ and  $N_{\rm eff} = 2.99 \pm 0.17$.

\section{Discussion}
\label{sec:4}
We find a good agreement of our estimate of \yp\ with the previous ones (except for the ITG14 estimate), see Table\,\ref{tab:helium_dets} and Fig. \ref{fig:Y(eta)}. Here we discuss  systematic effects that could have caused the observed shift of the ITG14 result.

\subsection{Determination of \texorpdfstring{$y_{wm}$}{Lg}}
\label{sec:4.1.1}
\begin{figure*}
    \centering
    \includegraphics[width = 0.97\textwidth]{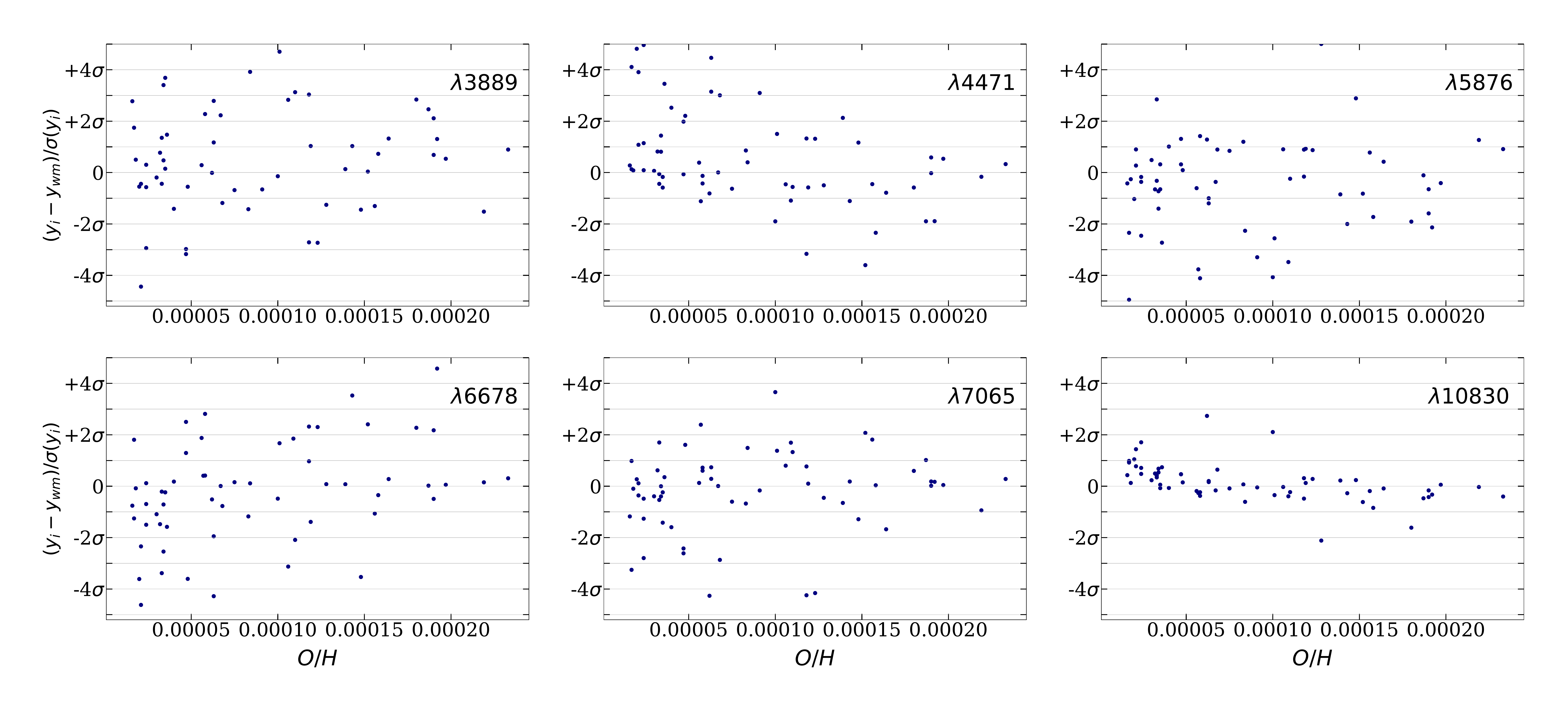}
    \caption{The distributions of $(y^+_i - y_{wm}) / \sigma(y^+_i)$ derived for six He lines: $\lambda 3889$, $\lambda 4471$, $\lambda 5876$, $\lambda 6678$, $\lambda 7065$, and $\lambda 10830$. All lines were used in the calculation of $y_{wm}$. }
    \label{fig:reldev}
\end{figure*}

The physical parameters of \HII\ regions in the ITG14 approach (the electron density $n_e$, electron temperature $T_e$, and optical depth $\tau_{3889}$ at $3889$\AA) are determined via the MCMC minimization of the following $\chi^2$ likelihood function: 
\begin{equation}
    \chi^2 = \sum_i \frac{(y_i^+ - y_{wm})^2}{\sigma(y_i^+)^2}.
	\label{eq:ITG14LH}
\end{equation}
where $y_i^+$ is the ratio of number densities of the single ionized helium and hydrogen derived for the He emission line labeled by $i$, $\sigma(y_i^+)$ is its uncertainty, $y_{wm}$ is a weighted mean, which defined as 
\begin{equation}
    y_{wm} = \frac{\sum_j \omega_j y^+_j }{\sum_j \omega_j}
\label{eq:ITG14LH2}
\end{equation}
where the statistical weight $\omega_j = 1 / \sigma(y^+_j)^2$. In the Eq.\,(\ref{eq:ITG14LH}) the summation over $i$ is carried out for six He lines: $\lambda 3889$, $\lambda 4471$, $\lambda 5876$, $\lambda 6678$, $\lambda 7065$, and $\lambda 10830$, while in the Eq.\,(\ref{eq:ITG14LH2}) the summation over $j$ is carried out only for four lines: $\lambda 4471$, $\lambda 5876$, $\lambda 6678$, and $\lambda 10830$. We suppose that this effect could introduce bias to the $y^+$ estimate. In the ITG14 approach the lines $\lambda 3889$ and $\lambda 7065$ are excluded from the $y_{wm}$ calculation since they show higher dispersion of the $y^+_j/ y_{wm}$ ratio around a value of 1 (see Fig. 3 in \citealt{ITG14}). However, this dispersion is not a correct characteristic of the ``goodness'' of the $j$ line, but it only demonstrates how the $j$ line affects the determination of $y_{wm}$. For a correct estimate of the ``goodness'' of the $j$ line, one have to consider the relative deviation $(y^+_j - y_{wm}) / \sigma(y^+_j)$ instead of the ratio $y^+_j / y_{wm}$. 

To check this point we calculate the distributions of $(y^+_i - y_{wm}) / \sigma(y^+_i)$ for all HeBCD objects. As can be seen from Fig. \ref{fig:reldev} all of the distributions have a comparable dispersion with only $\lambda10830$ line standing out. This is due to the fact that the NIR line has a significantly lower error of flux measurement compared to the optical lines. Therefore, we argue that there is no statistical reasons to exclude $\lambda 3889$ and $\lambda 7065$ lines from the $y_{wm}$ calculation. This exclusion in the ITG14 analyses might bias the $Y$ estimate.

\subsection{Helium underlying stellar absorption \texorpdfstring{$a_{\rm He}$}{Lg}}
The value of underlying absorption of the He lines is given in terms of the equivalent width and normalized to the value $a_{\rm He}$ for the $\lambda 4471$ line. In ITG14 the $a_{\rm He}$ value is fixed to be 0.4~\AA~ (while real a$_{\rm He}$ varies from 0 to 1~\AA. The $a_{\rm He}$ for other He lines are recalculated using the coefficients presented in \cite{ITG14}. Since the real value of underlying absorption can differ from 0.4~\AA, it can shift the intrinsic line fluxes. Therefore (similar to the AOS15 approach) we introduce $a_{\rm He}$ as a free parameter in the equation \ref{eq:ITG14LH}.

\subsection{\texorpdfstring{$T_e$}{Lg} determination}
\label{sec:4.1.3}
In ITG14 electron temperature was randomly varied within the range $(0.95 \times \widetilde{T_e}, 1.05 \times \widetilde{T_e})$ where 
\begin{equation}
    \widetilde{T_e} = T_e(\ion{O}{iii}) \times (2.51 \times 10^{-6}T_e(\ion{O}{iii}) + 0.8756 + 1152/T_e(\ion{O}{iii}))
    \label{eq:ITGtempprior}
\end{equation}
here the electron temperature $T_e(\ion{O}{iii})$ was derived from the ratio of [$\ion{O}{iii}$] emission lines fluxes $\lambda4363/(\lambda 4959 + \lambda5007)$.
This strict prior artificially over-constrains the range where MCMC routine searches for the best-fit value of $T_e$ . It leads to the concentration of the determined electron temperatures on lower or upper bounds of the prior. This effect was noted by \cite{ITG14} (see Fig. 4b and 4d therein). We suggest to remove this prior constraint as it can directly bias the value of $y_{wm}$. 

\subsection{ITG14 correction}
\label{sec:ITG14corr}
\begin{table}
	\centering
	\caption{Regression analyses for HeBCD and HeBCD + $S_f$ samples processed with different methods. The first column: the method used for the analysis, the second and the third columns: the results for the corresponding sample. }
	\label{tab:compar}
	\begin{tabular}{lcc} % four columns, alignment for each
		\hline
		Method & HeBCD & HeBCD + $S_{\rm f}$ \\
		\hline
		This paper & 0.2464 $\pm$ 0.0030 & 0.2462 $\pm$ 0.0022 \\
		ITG14 & 0.2528 $\pm$ 0.0023 & 0.2548 $\pm$ 0.0016 \\
		ITG14$_{\rm corr}$ & 0.2430 $\pm$ 0.0032 & 0.2449 $\pm$ 0.0020 \\
		\hline
	\end{tabular}
	\label{tab:itgcorrection}
\end{table}

The baseline of ITG14 method is presented in Appendix B. Applying ITG14 to HeBCD database we obtain the following estimate of $Y_p = 0.2528 \pm 0.0023$ which is consistent with the result of \cite{ITG14} $Y_p = 0.2551 \pm 0.0022$ (a slight deviation could have been caused by different sampling routines). Additionally, we apply ITG14 to the combination of $S_f$ and HeBCD  samples and obtain $Y_p = 0.2548 \pm 0.0016$. This estimate as well as the estimate by \cite{ITG14} exceeds all other estimates (see Tab. \ref{tab:helium_dets}). To assess the impact of the discussed effects (subsections \ref{sec:4.1.1} - \ref{sec:4.1.3}) on the estimate of $Y_p$, we have corrected the ITG14 procedure and applied it to the HeBCD and $S_f$ samples. We obtain the following results: $Y_p = 0.2430 \pm 0.0032$ for HeBCD and $Y_p = 0.2449 \pm 0.0020$ for HeBCD + $S_f$. It can be seen that after applying the discussed corrections to the ITG14 procedure, the estimate of $Y_p$ becomes consistent with other estimates (see Fig. \ref{fig:example_figure}). The results of the calculations are summarized in Table \ref{tab:itgcorrection}.

\section{Conclusions}
\label{sec:concl}

\begin{figure}
    %\centering
	\includegraphics[width=\columnwidth]{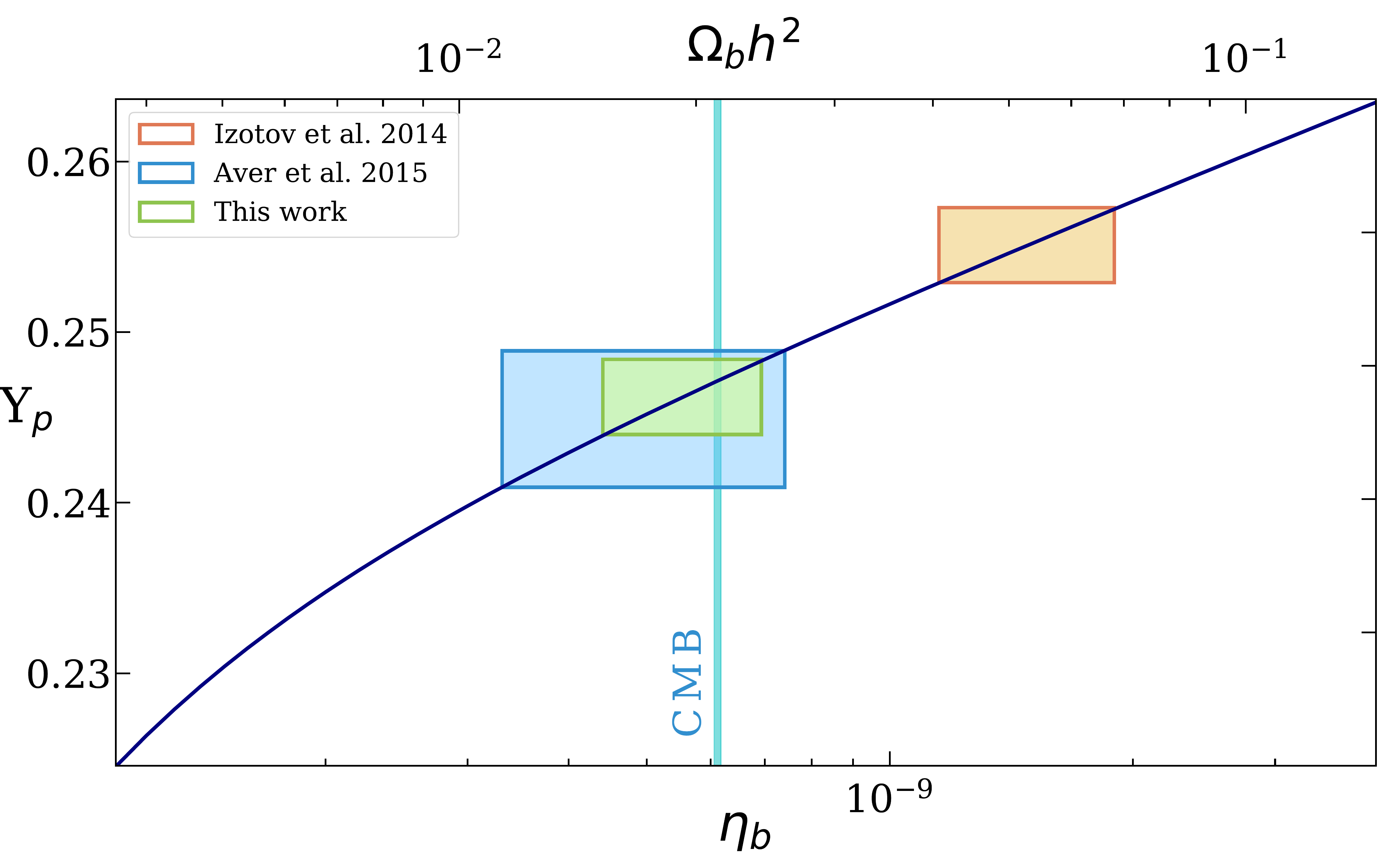}
    \caption{The primordial abundance of $^4$He as a function of the baryon-to-photon ratio $\eta_b$. The navy curve represents the calculation of Primordial Nucleosyntesis with our code \protect\citep{Orlov2000}. Orange, blue and green boxes represent the estimates of $Y_p$ obtained by \protect\cite{ITG14}, \protect\cite{AOS15}, and in this work. The vertical cyan line is the measurement of the baryon density based on the analysis of the CMB anisotropy \protect\citep{planck2018}.}
    \label{fig:Y(eta)}
\end{figure}

We scan the SDSS DR15 catalog for spectra of \HII regions in blue compact dwarf galaxies (BCD). Such objects are marked as the {\it{``STARBURST''}} in the SDSS catalog. We choose objects with the observational signal-to-noise ratio $S/N$ higher than 10. In total, we analysed 580 such objects. After processing the spectra with the approach similar to \cite{AOS15} we apply the $\chi^2$ selection criteria to the database. 100 objects satisfy the criteria and make up the final sample (Tab. \ref{tab:result_table}). Using this sample in combine with the HeBCD+NIR sample (20 objects) from \cite{ITG14} we report a new estimate of the primordial helium abundance $Y_p = 0.2462 \pm 0.0022$. The comparison of our result with the results from \cite{ITG14} and \cite{AOS15} are presented on Fig. \ref{fig:Y(eta)}.

We obtain the slope of $Y - {\rm O/H}$ relation $dY/d({\rm O/H}) = 46 \pm 13$. This slope is determined on $3.5 \sigma$ confidence level which is significantly higher compared to the previous studies. 

Using our value of $Y_p$ and the primordial deuterium abundance D/H from \cite{10.1093/ptep/ptaa104} we constrain the effective number of neutrino species $N_{\rm eff} = 2.95 \pm 0.16$ and baryon to photon ratio $\eta_{10} = 6.14 \pm 0.09$. The results are in good agreement with the Planck results of $\eta_{10} = 6.13 \pm 0.04$ and $N_{\rm eff} = 2.99 \pm 0.17$.

As a result of the detailed consideration of \cite{ITG14} method of $Y_p$ determination we suppose that we have found the reason for $Y_p$ overestimation.

\subsubsection*{Further improvements}
The SDSS catalog contains a large number of \HII region spectra in BCDs. We show that the data can be used to increase the regression statistics and thus significantly improve the accuracy of the estimate of $Y_p$. Unfortunately, manual data processing is difficult and time-consuming due to a large amount of spectral data. We plan to develop an automated procedure of SDSS spectra processing in order to significantly increase the statistics. Further increase of statistics potentially allows us to achieve Planck accuracy, which in turn will become a powerful tool for studying the self-consistency of the Standard Cosmological Model and/or physics beyond.

Beyond further enlarging of the data set with SDSS observations, the accuracy of the determination of \yp\, can be improved by adding more high quality spectra to the analysis and by refining spectra processing. High quality spectra can be obtained by observations of certain SDSS objects, for instance, from the final sample $S_{\rm f}$. In addition, high quality spectra can allow us to use additional weak H and He lines in the analysis, it will increase the reliability of determining the physical parameters of \HII regions. The spectra processing improvement could include measurements of line fluxes separate from underlying absorption features. To account for the underlying absorption two additional parameters are included into the photoinization model. The direct measurement of the underlying absorption features will remove the associated parameters from the model, that in turn may increase the accuracy of determining other parameters.

\section*{Data availability}

The observational data used in this paper is available from the public data archives of SDSS (SDSS SkyServer). 

\section*{Acknowledgements}
The authors thank the anonymous referee for useful suggestions. The authors are grateful to P. Shternin for discussions on statistical methods of data analysis, and E. Aver for useful comments and providing data on He emissivities. The work was supported by the Russian Science Foundation (grant 18-12-00301).

%%%%%%%%%%%%%%%%%%%%%%%%%%%%%%%%%%%%%%%%%%%%%%%%%%



\bibliographystyle{mnras}
\bibliography{Pr_Helium} % if your bibtex file is called example.bib

\appendix

\section{Photoinization model}
\label{sec:app:a}

Here we describe our photoionization model of the \HII\ region. We use the approach similar to one described by \cite{AOS15}.

The helium mass fraction $(Y=m_{\rm He}/m_{\rm gas})$ of the \HII\ region is derived by the analytic expression %
\begin{equation}
    Y = \frac{4y}{1+4y}(1-Z)
\end{equation}
where $y$ is the ratio of the total number densities of helium and hydrogen
\begin{equation}
y=\frac{n_{\rm He}}{n_{\rm H}},
\end{equation}
$Z$ is the total metallicity, which is connected with the total oxygen abundance (O/H) with the following equation (\cite{aver2010}):
\begin{equation}
Z = 20 \times {\rm{O/H}}.
\end{equation}

The helium abundance is given by a sum of abundances of its different ionization states:
\begin{equation}
    y = y^0 + y^+ + y^{++} = ICF\times\left(y^+ + y^{++}\right)
\end{equation}
here $y^0$, $y^+$ and $y^{++}$ are abundances of neutral, single and double ionized helium, the $ICF$ is the ionization correction factor accounting a part of neutral helium. We assumed this contribution to be negligible and set the $ICF=1$.

To estimate parameters $y^+$ and $y^{++}$ we fit the observed fluxes of the helium and hydrogen emission lines using analytic functions. We determine ${\rm O/H}$ using the direct method from \cite{izotov2006}.

The relative fluxes of helium and hydrogen recombination lines are given by 
\begin{equation}
\label{eq:AOS15he}
    \begin{split}
    &\frac{F_{\rm He}(\lambda)}{F({\rm H}\beta)}_{\rm theor} = y^+ \times\frac{E_{\rm He}(\lambda)}{E({\rm H}\beta)}  \times \frac{f_\tau(\lambda)}{1 + \frac{C}{R}({\rm H}\beta)} \times\\
    &\times\frac{EW({\rm H}\beta) + a_{\rm H}({\rm H}\beta)}{EW({\rm H}\beta)} \frac{EW(\lambda)}{EW(\lambda) + a_{\rm He}(\lambda)}\times10^{-f(\lambda) C({\rm H}\beta)}
    \end{split}
\end{equation}
and 
\begin{equation}
    \label{eq:AOS15h}
    \begin{split}
    &\frac{F_{\rm H}(\lambda)}{F({\rm H}\beta)}_{\rm theor} = \frac{E_{\rm H}(\lambda)}{E({\rm H}\beta)}\times \frac{1 + \frac{C}{R}(\lambda)}{1 + \frac{C}{R}({\rm H}\beta)} \times \\
    & \times\frac{EW(H\beta) + a_{\rm H}({\rm H}\beta)}{EW({\rm H}\beta)} \frac{EW(\lambda)}{EW(\lambda) + a_{\rm H}(\lambda)}\times10^{-f(\lambda) C({\rm H}\beta)}  
    \end{split}
\end{equation}
Here $E_{\rm He, H}(\lambda)$ are the helium and hydrogen  emissivity functions of $n_e$ and $t_e$, $f_\tau(\lambda)$ is optical depth function, $\frac{C}{R}(\lambda)$ is the correction for collision excitation of hydrogen lines, $f(\lambda)$ is the correction for interstellar reddening. The factors containing EWs and $a_{\rm He, H}(\lambda)$ are responsible for the helium and hydrogen underlying stellar absorption. These terms are discussed in details below. 

The helium emissivity $E_{\rm He}(\lambda)$ is calculated using the bilinear interpolation in the fine grid of the electron density and temperature presented by \cite{aver2013}. %The authors calculate emissivities on a grid of temperatures varying within 10000 $\div$ 22000 K and for densities within 0 $\div$ 10000 cm$^{-3}$. The helium emissivities are calculated by bilinear interpolation of the grid.
The hydrogen emissivity $E_{\rm H}(\lambda)$ is calculated using data from \cite{1987MNRAS.224..801H}. The emissivity of H($\beta$) line is calculated using the following expression:
\begin{equation}
E({\rm H}\beta) = \left[a - b (\ln (t_e))^2 + c \ln (t_e) + \frac{d}{\ln(t_e)}\right]\times{t_e^{-1}},
\end{equation}
where $a = -8.19744 \times 10^5$, $b = 3431.6$, $c = 93354.4$, $d = 2.425776 \times 10^6$. The emissivity of other hydrogen lines is given by 
\begin{equation}
\label{eq:hedemis}
    E_{\rm H}(\lambda) = \sum_{ij} A_{ij} (\log_{10} (t_e))^i (\log_{10}(n_e))^j
\end{equation}
Coefficients $A_{ij}$ are presented in Tab.\,\ref{tab:HydEm}. 

\begin{table}
	\centering
	\caption{Coefficients for the hydrogen emissivities $A_{ij}$.}
	\label{tab:HydEm}
	\begin{tabular}{ccccc} % four columns, alignment for each
		\hline
		Line & i$\downarrow$ & j$\rightarrow$ \\
		\hline
		 & & 0 & 1 & 2 \\
		 \hline
		${\rm H}\alpha$ & 0 & 2.8339 & 0.0221 & $-$0.0045\\
		& 1 & 0.4120 & $-$0.6390 &  0.1065\\
		& 2 & $-$1.7715 & 1.5136 & -0.2523\\
		\hline
		${\rm H}\gamma$ & 0 & 0.4635 & 0.0031 & $-$0.0004 \\
		& 1 & 0.1227 & $-$0.0633 & 0.0098 \\
		& 2 & $-$0.3401 & 0.2186 & -0.0342\\
		\hline
		${\rm H}\delta$ & 0 & 0.2626 & $-$0.0029 & 0.0005\\
		& 1 & $-$0.0567 & 0.0575 & $-$0.0097\\
		& 2 & 0.1990 & $-$0.1565 & 0.0260\\
		\hline
		${\rm P}\gamma$ & 0 & 0.0904 & 0.0001 & $-$0.00004\\
		& 1 & $-$0.0274 & $-$0.0005 & 0.0001\\
		& 2 & 0.0040 & $-$0.0004 & 0.0001\\
		\hline
		${\rm H}8$ & 0 & 0.1058 & $-$0.0005 & 0.00007\\
		& 1 & $-$0.0370 & 0.0309 & $-$0.0044\\
		& 2 & 0.1306 & $-$0.0934 & 0.0133\\
		\hline
	\end{tabular}
\end{table}

Following  \cite{aver2010} we now define the $\frac{C}{R}(\lambda)$ factor 
\begin{equation}
\frac{C}{R}(\lambda) = \xi \sum_i a_i(\lambda) \exp \left(-\frac{b_i(\lambda)}{t_e} \right) t_e ^{c_i(\lambda)}
\end{equation}
where $\xi=n_{\rm H}/n_{\rm H^+}$ is the ratio of densities of neutral and ionized hydrogen, $a_i$, $b_i$ and $c_i$ are coefficients presented in Tab.\,\ref{tab:hydCR}. 

\begin{table*}
	\centering
	\caption{Coefficients for the hydrogen collisional to recombination correction $C/R(\lambda)$.}
	\label{tab:hydCR}
	\begin{tabular}{ccccccccccc} % four columns, alignment for each
		 \hline
		${\rm H}\alpha$ & a & 0.4155 & 2.4965 & 2.4063 & 0.2914 & 0.3685 & 4.6426 \\
		& b & 14.80 & 14.30 & 14.03 & 14.80 & 14.80 & 14.03 \\
		& c & 0.4209 & 0.5853 & 0.6187 & 0.6766 & 0.7076 & 0.7788 \\
		\hline
		${\rm H}\beta$ & a & 0.2384 & 0.6964 & 0.1991 & 0.1409 & 0.2201 & 1.9228 & 1.4845 & 2.8179 \\
		& b & 15.15 & 14.80 & 15.15 & 15.15 & 15.15 & 14.80 & 14.80 & 14.80  \\
		& c & 0.3082 & 0.4978 & 0.6017 & 0.6765 & 0.7293 & 0.7535 & 0.7845 & 0.9352 \\
		\hline
		${\rm H}\gamma$ & a & 0.3629 & 0.8351 & 2.0044 & 1.4757 & 2.7947 \\
		& b & 15.15 & 15.15 & 15.15 & 15.15& 15.15 \\
		& c & 0.3598 & 0.6533 & 0.7281 & 0.7809 & 0.8582\\
		\hline
		${\rm H}\delta$ & a & 0.3629 & 0.8351 & 2.0044 & 1.4757 & 2.7947 \\
        & b & 15.34 & 15.34 & 15.34 & 15.34 & 15.34 \\
		& c & 0.3598 & 0.6533 & 0.7281 & 0.7809 & 0.8582\\
		\hline
	\end{tabular}
\end{table*}

Then optical depth function $f_\tau$ is used to make a correction for photons that are reabsorbed or scattered out inside the \HII\ region. The corrections for each helium line is calculated individually using expression from \cite{benjamin2002}:
\begin{equation}
    f_\tau(\lambda) = 1 + \frac{\tau}{2} (a + t_e \times (b_0 + b_1 n_e + b_2 n_e^2))
\end{equation}
Coefficients $a$, $b_0$, $b_1$ and $b_2$ are presented in Tab. \ref{tab:odf}.

\begin{table}
	\centering
	\caption{Coefficients of the optical depth function.}
	\label{tab:odf}
	\begin{tabular}{ccccc} % four columns, alignment for each
		\hline
		Line & a & $b_0$ & $b_1$ & $b_2$ \\
		\hline
		3889 & $-$1.06 $\times 10^{-1}$ & 5.14 $\times 10^{-5}$ & $-$4.20 $\times 10^{-7}$ & 1.97 $\times 10^{-10}$\\
        4026 & 1.43 $\times 10^{-3}$ & 4.05 $\times 10^{-4}$ & 3.63 $\times 10^{-8}$ & ...\\
        4471 & 2.74 $\times 10^{-3}$ & 0.81 $\times 10^{-4}$ & $-$1.21 $\times 10^{-6}$ & ... \\
        5876 & 4.70 $\times 10^{-3}$ & 2.23 $\times 10^{-3}$ & $-$2.51 $\times 10^{-6}$ & ... \\
        6678 & 0 & 0 & 0 & 0  \\
        7065 & 3.59 $\times 10^{-1}$ & $-$3.46 $\times 10^{-2}$ & $-$1.84 $\times 10^{-4}$ & 3.039 $\times 10^{-7}$ \\
        10830 & 1.49 $\times 10^{-2}$ & 4.45 $\times 10^{-3}$ & $-$6.34 $\times 10^{-5}$ & 9.20 $\times 10^{-8}$ \\
		\hline
	\end{tabular}
\end{table}

The observed helium and hydrogen line fluxes are also affected by the underlying stellar absorption and interstellar reddening (third factor in Eq.\,\ref{eq:AOS15he} and Eq.\,\ref{eq:AOS15h}). First, the underlying absorption parameters can be expressed as:
\begin{equation}
    \begin{split}
    &a_{\rm He}(\lambda) = A(\lambda) \times a_{\rm He}(4471)\\
    &a_{\rm H}(\lambda) = B(\lambda) \times a_{\rm H}(H\beta)  
    \end{split}
\end{equation}
where coefficients $A(\lambda)$ and $B(\lambda)$ presented in Tab. \ref{tab:AOSua}. Then, the correction of the interstellar reddening can be accounted by using of the combination of the logarithmic correction factor $C({\rm H}\beta)$ and the reddening function $f(\lambda)$, which are described in \cite{1989ApJ...345..245C}.

\begin{table}
	\centering
	\caption{Wavelength dependence coefficients for the underlying helium and hydrogen absorption.}
	\label{tab:AOSua}
	\begin{tabular}{cc|cc} % four columns, alignment for each
		\hline
        Line & $A(\lambda)$ & Line & $B(\lambda)$ \\
        \hline
        $\lambda3889$ & 1.400 & H8 & 0.882 \\
        $\lambda2046$ & 1.347 & H$\gamma$ & 0.896 \\
        $\lambda4471$ & 1.000 & H$\delta$ & 0.959 \\
        $\lambda5876$ & 0.874 & H$\beta$ & 1.000 \\
        $\lambda6678$ & 0.525 & H$\alpha$ & 0.942 \\
        $\lambda7065$ & 0.400 & P$\gamma$ & 0.400 \\
        $\lambda10830$ & 0.400 & & \\
		\hline
	\end{tabular}
\end{table}

Secondly, we take into account that the He$\lambda3889$ line is blended with the H8 emission line. We separate them by subtracting the flux of ${\rm H}8$ line calculated with the equation \ref{eq:hedemis}. Therefore we now may write:
\begin{equation}
    \begin{split}
    &\frac{F({\rm He}\lambda3889)}{F({\rm H}\beta)} = \frac{F({\rm H}8 + {\rm He}\lambda3889)}{F({\rm H}\beta)}\frac{EW({\rm H}8) + a_{\rm H}({\rm H}8)}{EW({\rm H}8)} - \\
    &\frac{E_({\rm H}8)}{E({\rm H}\beta)} 10^{-f({\rm H}8) C({\rm H}\beta)}
    \end{split}
\end{equation}

Therefore our model have 8 fitting parameters (the abundance of the single ionized helium $y^+$, electron density $n_e$, electron temperature $t_e$, optical depth $\tau_{3889}$, underlying stellar H and He absorption parameters $a_H$ and $a_{He}$, reddening parameter $C(H\beta)$ and the fraction of neutral hydrogen $\xi$)  which are
determined by minimizing of the likelihood function 
\begin{equation}
    \chi^2 = \sum_i \frac{\left(\frac{F(\lambda_i)}{F(H\beta)}_{\rm theor} - \frac{F(\lambda_i)}{F(H\beta)}_{\rm obs}\right)^2}{\sigma_{obs}(\lambda_i)^2}
    \label{eq: LF}
\end{equation}
where the summation is over the sample of 7 helium  ($\lambda$3889, $\lambda$4026, $\lambda$4471, $\lambda$5876, $\lambda$6678, $\lambda$7065 and $\lambda$10830) and 3 hydrogen lines (H$\alpha$, H$\gamma$, H$\delta$). 
The parameters are varied in the range of $0.00 < y^+ < 0.15$, $0 < n_e < 2000$, $1.0 < t_e < 2.2$, $0.0 < a_{He} < 3.0$, $0.0 < a_{H} < 8.0$, $0.0 < \tau < 5.0$, $0.00 < C(H\beta) < 0.99$, $0 < \xi < 3000$. We use the best-fit values of $t_e$ and C(H$\beta)$ to estimate the value of $y^{++}$ with the following equation:

\begin{equation}
    y^{++} = 0.084t_e^{0.14} \frac{F(\lambda4686)}{F(\rm H \beta)} 10^{{\rm C(H}\beta) f(\lambda4686)}
\end{equation}

Following \cite{aver2011} we add the prior distribution for the electron temperature:
\begin{equation}
    \label{eq:prior}
    \chi^2_T = \frac{(t_e - t(\text{O\,{\sc iii}}))^2}{(0.2 t(\text{O\,{\sc iii}}))^2}
\end{equation}
where $t$({O\,{\sc iii}}) is the electron temperature derived from the analysis of [O\,{\sc iii}] emission line fluxes.

Following \cite{izotov2006} we determine the temperature $t(\text{O\,{\sc iii}})) = 10^{-4} \times T$(O\,{\sc iii}) using:
\begin{equation}
\label{eq:toiii}
    t = \frac{1.432}{\log_{10}\Big( \frac{\lambda4959 + \lambda5007}{\lambda4363}\Big) - \log_{10}C_T}
\end{equation}

Here and after letter $\lambda$ with the specific wavelength denotes the measured flux of corresponding ion line normalized to the H$\beta$ flux. The $C_T$ term is given by
\begin{equation}
C_T = (8.44 - 1.09t + 0.5t^2 - 0.08t^3 ) \frac{1 + 0.0004x}{1 + 0.044x}
\end{equation}
where $x = 10^{-4}n_e\sqrt{t}$. 

Temperature $t$ from \ref{eq:toiii} is determined via iterative process starting with $t = 1.0$. Such iteration gives correct results after 6 iteration steps. To determine $n_e$ needed for the $C_T$ term calculation we use fitting formula from \cite{2014A&A...561A..10P}:
\begin{equation}
\label{eq:eden}
\begin{split}
    &\log_{10}(n_e) = 0.0543 \times \tan(-3.0553 R + 2.8506) + 6.98 - \\
    & - 10.6905 R + 9.9186 R^2 - 3.5442R^3
\end{split}
\end{equation}

here $R = \lambda6717 / \lambda6731$. The fitting formula \ref{eq:eden} is derived for $T$(S\,{\sc ii}) = 10$^4$ K. The electron density for different $T$(S\,{\sc ii}) can be calculated with following scaling factor:
\begin{equation}
    n_e(T(\text{S\,{\sc ii}})) = n_e(10^4K) \times \sqrt{\frac{T(\text{S\,{\sc ii}})}{10^4K}}
\end{equation}

According to \cite{izotov2006} we consider $T(\text{S\,{\sc ii}}) = T(\text{O\,{\sc ii}})$ which is calculated using following relation:
\begin{equation}
    \begin{split}
         &T(\text{O\,{\sc ii}}) = -0.577 + t \times (2.065 - 0.498t)~,~~\text{A(O/H) < 7.2}\\
         &T(\text{O\,{\sc ii}}) = -0.744 + t \times (2.338 - 0.610t)~,~~\text{7.2 $\leq$ A(O/H) $\leq$ 7.6}\\
         &T(\text{O\,{\sc ii}}) = 2.967 + t \times (-4.797 + 2.827t)~,~~\text{A(O/H) > 7.6}
    \end{split}
\end{equation}
here $A({\rm O/H}) = 12 + \log_{10}({\rm O/H})$. 

The oxygen abundance O/H is determined by expression:
\begin{equation}
    \frac{\rm O}{\rm H} = \frac{\rm O^+}{\rm H} + \frac{\rm O^{++}}{\rm H}
\end{equation}

The ionic abundances are determined using following formulas from \cite{izotov2006}:
\begin{equation}
\begin{split}
        &12 + \log_{10}\Big(\frac{\rm O^{++}}{\rm H}\Big) = \log_{10}\Big( \frac{\lambda4959 +  \lambda5007}{\lambda4363}\Big) + 6.2 +\frac{1.251}{t} -\\
        & -0.55\log_{10}t -0.14t
\end{split}
\end{equation}
\begin{equation}
\begin{split}
        &12 + \log_{10}\Big(\frac{\rm O^{+}}{\rm H}\Big) = \log_{10}(\lambda7320 + \lambda 7330 ) + 6.901 +\frac{2.487}{t} -\\
        & -0.483\log_{10}t -0.013t + \log_{10}(1 - 3.48x)
\end{split}
\end{equation}

\section{Photoinization model ITG14}
\label{sec:app:b}

Here we describe the ITG14 method of the determination of physical properties of observed \HII\ regions in blue compact dwarf galaxies. The crucial difference with the AOS15 self-consistent approach is that ITG14 involves a step by step correction for the systematic effects.

The determination of $Y_p$ with the \cite{ITG14} method involves calculation of current helium and oxygen abundances, $Y$ and O/H.

The current helium mass fraction $Y$ is determined via same equation as presented in \ref{sec:app:a}:
\begin{equation}
    Y = \frac{4y}{1 + 4y} (1 - Z)
\end{equation}
Here $y$ is total helium to hydrogen abundance ratio and $Z = B \times {\rm O/H}$. Unlike \cite{AOS15}, where the authors set $B = 20$, \cite{ITG14} determine $B$ using following relation:
\begin{equation}
    B = 8.64(12 + \log ({\rm O/H})) - 47.44
\end{equation}
$y$ is determined via the following equation:
\begin{equation}
    y = ICF({\rm He}) (y^+ + y^{++})
\end{equation}

The following steps of systematic effects corrections involve the determined O/H. The oxygen abundance alongside the temperature $T$(O\,{\sc iii}) and the electron density $n_e$ is determined in the same way presented in \ref{sec:app:a}.

The method is focused in the pure recombination value of hydrogen intensity, and thus hydrogen line intensities should be corrected for collisional and fluorescent excitation. This is done with the coefficient $(C+F)/I$, where $C$ and $F$ are collisional and fluorescent contributions to the intensity. This quantity is determined via equations (16 - 27) from \cite{izotov2013} as a function of oxygen abundance. 

The fluxes are corrected for both interstellar extinction and hydrogen underlying absorption using an interative procedure from \cite{1994ApJ...435..647I}. Reddening parameter $C({\rm H}\beta$) and underlying absorption parameter $a_{\rm H}$ are determined via comparison of observed and theoretical Balmer decrement values with the following equations:
\begin{equation}
    \frac{I(\lambda)}{I({\rm H}\beta)} = \frac{F(\lambda)}{F({\rm H}\beta)} \frac{EW(\lambda) + a_{\rm H}}{EW(\lambda)} \frac{EW({\rm H}\beta)}{EW({\rm H}\beta) + a_{\rm H}}  10 ^{f(\lambda) C({\rm H}\beta)}
\end{equation}
Here $I(\lambda)$ and $F(\lambda)$ denote intrinsic and observed line flux respectively, $EW$ is the equivalent width of a line, $f(\lambda)$ is the reddening function from \cite{1989ApJ...345..245C}. It is assumed that $a_{\rm H}$ is the same for all hydrogen lines. The intrinsic Balmer decrement is calculated using \cite{1987MNRAS.224..801H}. After this parameters are determined the whole observed spectrum is corrected for these effects. 

The He line fluxes are corrected for underlying stellar absorption. This is done using the following equation:
\begin{equation}
    \frac{I_{\rm He}(\lambda)}{I({\rm H}\beta)} = \frac{F_{\rm He}(\lambda)}{F({\rm H}\beta)} \frac{EW_{\rm He}(\lambda) + a_{\rm He}(\lambda)}{EW_{\rm He}(\lambda)}
\end{equation}
The absorption line equivalent width $a_{\rm He}(4471)$ is fixed on 0.4~\AA. The equivalent widths of the other absorption lines were fixed according to the ratios (one can compare them with coefficients presented below:
\begin{equation}
    \begin{split}
        &a_{\rm He}(3889) / a_{\rm He}(4471) = 1.0 \\
        &a_{\rm He}(5876) / a_{\rm He}(4471) = 0.8 \\
        &a_{\rm He}(6678) / a_{\rm He}(4471) = 0.4 \\
        &a_{\rm He}(7065) / a_{\rm He}(4471) = 0.4 \\
        &a_{\rm He}(10830) / a_{\rm He}(4471) = 0.8 
    \end{split}
\end{equation}

Determination of single ionized helium to hydrogen abundance ratio is determined via minimization of the quantity:
\begin{equation}
    \chi^2 = \sum_i \frac{(y_i^+ - y_{wm})^2}{\sigma(y_i^+)^2}
	\label{eq:appendBLH}
\end{equation}
Here $y_i^+$ is a single ionized He number density derived from the intensity of the $i$ He emission line (see eq. \ref{appendyp}), $\sigma(y_i^+)$ is the uncertainty propagated from the measured error of $i$-line flux, $y_{wm}$ is the weighted mean of $y^+_i$. The summation over $i$ is carried for the following He lines: $\lambda 3889$, $\lambda 4471$, $\lambda 5876$, $\lambda 6678$, $\lambda 7065$, and $\lambda 10830$, but only lines $\lambda 4471$, $\lambda 5876$, $\lambda 6678$, and $\lambda 10830$ are used in $y_{wm}$ calculation. 

\begin{equation}
\label{appendyp}
    y_i^+ = R_i \frac{E_{{\rm H}\beta}(n_e, T_e)}{E_{i}(n_e, T_e) f_i(n_e, T_e, \tau)}
\end{equation}
Here $R_i$ is the $i$ He line flux corrected for reddening, underlying H and He absorption and hydrogen non-recombination contribution, $E_{{\rm H}\beta}(n_e, T_e)$ is the emissivity of H$\beta$ line taken from \cite{1987MNRAS.224..801H}, $E_{i}(n_e, T_e)$ is the $i$ He line emissivity calculated using analytical fits from \cite{izotov2013}. The optical depth function $f_i(n_e, T_e, \tau)$ is taken from \cite{benjamin2002}.

The likelihood function \ref{eq:appendBLH} is minimized over $n_e$, $T_e$, and $\tau$ which are varied within the following ranges:
\begin{equation}
    \begin{split}
        &0 < n_e < 600 \\
        &0.95~ T_{pr} < T_e < 1.05~ T_{pr} \\
        &0 < \tau < 5
    \end{split}
\end{equation}
here $T_{pr}$ is derived using the following relation from \cite{izotov2013}:
\begin{equation}
    T_{pr} = \big( 2.51 \times 10^{-6}T_e (\text{O\,{\sc iii}}) + 0.8756 + 1152 / T_e (\text{O\,{\sc iii}}) \Big) \times T_e (\text{O\,{\sc iii}})
\end{equation}

Finally $y_{wm}$ is determined using the best-fit parameters $T_e$, $n_e$ and $\tau$. In case the He\,{\sc ii} $\lambda 4686$ is detected, $y^{++}$ is calculated.

The total helium abundance $y$ is determined with the following equation:
\begin{equation}
    y = ICF ({\rm He} ) \times (y_{wm} + y^{++})
\end{equation}

Unlike AOS15, ITG14 use the ionization correction factor $ICF ({\rm He})$ which is calculated as a function of oxygen excitation parameter $x =$ O$^{2+} / ({\rm O^+ + O^{2+}})$. The analytical fits for this quantity are presented in \cite{izotov2013}.

%%%%%%%%%%%%%%%%%%%%%%%%%%%%%%%%%%%%%%%%%%%%%%%%%%

% Don't change these lines
\bsp	% typesetting comment
\label{lastpage}
\end{document}